\documentclass[twocolumn,aps,pra,showpacs]{revtex4}

\usepackage{epsfig}
\usepackage{graphics}

\newcommand{\ha}{\hat{a}}

\newcommand{\hc}{\hat{c}}

\begin{document}

%\preprint{}

\title{On-demand generation of entanglement of atomic qubits via optical interferometry}


%\author{}
%\thanks{}

\author{Y. P.  Huang}
\author{M. G. Moore}
\affiliation{Department of Physics and Astronomy, Michigan State
University, East Lansing, MI 48824}


%\date{\today}

\begin{abstract}
The problem of on-demand generation of entanglement between
single-atom qubits via a common photonic channel is examined within
the framework of optical interferometry. As expected, for a
Mach-Zehnder interferometer with coherent laser beam as input, a
high-finesse optical cavity is required to overcome sensitivity to
spontaneous emission. We show, however, that with a twin-Fock input,
useful entanglement can in principle be created without
cavity-enhancement. Both approaches require single-photon resolving
detectors, and best results would be obtained by combining both
cavity-feedback and twin-Fock inputs. Such an approach may allow a
fidelity of $.99$ using a two-photon input and currently available
mirror and detector technology. In addition, we study
interferometers based on NOON states and show that they perform
similarly to the twin-Fock states, yet without the need for
high-precision photo-detectors. The present interferometrical
approach can serve as a universal, scalable circuit element for
quantum information processing, from which fast quantum gates,
deterministic teleportation, entanglement swapping $etc.$, can be
realized with the aid of single-qubit operations.

\end{abstract}

\maketitle

\section{Introduction}
\label{introduction} Practical quantum information processing will
rely on deterministic computational gates and high-fidelity
communication protocols that operate successfully on-demand
\cite{NieChu00,BenBraCre93,BouPanMat97}. This requires realtime
generation of entanglement amongst arbitrary qubits performed at
near-unit success probability and fidelity. For atom-type qubits,
this entanglement can be generated either via a photonic channel,
utilizing entangled photon-pairs \cite{BouPanMat97,MarRieTit03} or
cavity-decay photons
\cite{BosKniPle99,ZheGuo06,DuaKim03,DiMutScu05}, or an atomic
channel as in recent trapped-ion experiments
\cite{RieHafRoo04,BarChiSch04}.  For high-speed quantum computation
and/or long-distance communication, a photonic quantum channel is
clearly ideal, as photons are robust carriers of quantum information
that travel at the speed of light. Since isolated trapped-atomic
qubits have long coherence times and are easily manipulated with
electromagnetic fields, it is of general interest to consider the
problem of creating entanglement between two isolated atomic qubits
via their mutual interaction with a single photonic channel. The
primary obstacle to such a protocol lies in the problem of
eliminating spontaneous emission while obtaining a sufficiently
strong atom-photon interaction. Recent attempts to overcome this
difficulty have primarily relied on the use of collective-state
qubits in atomic ensembles to enhance the dipole moment of the qubit
\cite{DuaCirZol00,JulKozPol01,MatKuz04}. This enhancement effect
has allowed Duan, Cirac, Zoller and Polzik to implement a quantum teleportation
scheme between two atomic samples, where a coherent beam is passed
successively through and the entanglement is generated by measuring
its final Faraday-rotation angle \cite{DuaCirZol00}. Very
recently, a probabilistic scheme to entangle two distant quantum
dots using cavity enhancement has been proposed using bright
coherent light via homodyne detection and postselection
\cite{LooLadSan06}.

In this paper, we investigate an approach in which {\it single}-atom
qubits are deterministically entangled by use of an optical
interferometer, thus avoiding collisional decoherence mechanisms
inherent in atomic ensembles. It is well-known that the back-action
of a single atom onto a focused laser pulse is very weak, so that
generating useful atom-photon entanglement in this manner will
generally fail due to spontaneous emission
 \cite{EnkKim01}. Our goal, however, is to overcome this difficulty by
using the extreme sensitivity of sub-shot-noise
interferometers
\cite{YurMcCKla86,HolBur93,PezSme06,HeiHorRey87,LanFriRon02,BolItaWin96,HueMacPel97,Ger00,MunNemMil02}
to detect the weak phase imprinted on the forward scattered light in
the regime where spontaneous emission is negligible. In addition, we
also consider the more generic approach of using high-finesse
optical resonators \cite{HooKimYe01, RaiBruHar01} to enhance the
atom-photon interaction. Our interferometry apparatus 
follows the Faraday-rotation scheme of Duan,  Cirac, Zoller, and Polzik
\cite{DuaCirZol00}, with the collective atomic ensembles replaced by single trapped atoms, and with the coherent light replaced by a highly non-classical many-photon state. We first show that for a Mach-Zehnder
(MZ) interferometer with coherent input, a high-finesse Q-switch
cavity is always necessary, and to achieve a fidelity of $f=.99$
requires an optical cavity which cycles photon for $M=10^5-10^6$
times. If the coherent state input is replaced with a twin-Fock (TF)
input state, however, we find that a cavity is in principle no
longer required. Cavity feedback may still provide additional
improvement in performance. For example, $f=.99$ can be achieved if
we use the TF state with $4\times 10^4$ photons and no cavity, or
only two photons and cavities with $M=2\times 10^4$. The later
requires a single photon-on-demand \cite{KuhHenRem02} injected into
each interferometer input, with an accurate measurement of the
two-photon output state, which appears within the realm of
experimental feasibility. Both MZ-interferometer-based approaches
require detectors with single-photon resolution \cite{IrvHenBou06}.
This requirement, however, can be overcome by employing a non-MZ
interferometer based on NOON states and nonlinear beamsplitters.
Such an interferometer yields a sensitivity close to the TF state in
detecting phase imbalance, and thus can achieve similar performance
without counting single photons. While the TF and NOON
states have recently been shown as unable to measure any phase below shot-noise in a 
single measurement \cite{HraReh05,PezSme05,PezSme06}, our present work shows that single-measurements with these
states can still be highly useful as `quantum switches' with
Heisenberg-limited sensitivity.

Our proposed interferometry approach to entangle atomic-qubits can
be performed on-demand and is scalable. We envision generalizing
such a device to a complete set of quantum information processing
protocols whereby stationary single-atom qubits are held in isolated
traps, with arbitrary single-atom and multi-atom operations achieved
via sequences of light pulses guided amongst the atoms and into
detectors by fast optical switching. The goal of this paper is to
perform a theoretical analysis of interferometrical generation of
entanglement between two arbitrary qubits, and to determine the
fundamental limitations imposed by quantum mechanics.

The paper is organized as follows. In Sec.\ref{model}, we present a
basic model of the interferometrical generation of entanglement
between two atomic qubits. In Sec.\ref{cohrt} and
Sec.\ref{twinfock}, we study two MZ-interferometrical approaches
using the coherent and the TF input light field, respectively. Then
in Sec.\ref{noon}, we investigate an alternative approach employing
NOON states and nonlinear beamsplitters. In Sec.\ref{app}, as
examples, we briefly show how the present scheme can be applied to
realize deterministic teleportation, multi-site entanglement, and
entanglement swapping. This is followed by a short discussian and
conclusion in Sec.\ref{conclusion}

\section{The model}
\label{model}
In our scheme, a single pulse of light is passed
through an optical interferometer, with the different 'arms' of the
interferometer corresponding to different photon polarization
states. The beam passes through two atomic qubits, i.e. trapped
ions, neutral atoms and/or quantum dots, such that each polarization
state interacts with a different internal atomic state. This can be
achieved using an 'X'-type scheme, as described in
\cite{DuaCirZol00}, in which the Zeeman sublevels of an $F=1/2$
ground state form the qubit, or in a $\Lambda$-type level scheme,
with the $m=\pm1$ states of an $F=1$ ground state forming the
qubits. In both cases, the 'arms' of the interferometer would
correspond to orthogonal circular polarization states.  The
interferometer output is determined by a state-dependent phase-shift
acquired via the atom-photon interaction. This requires a large
detuning from the atomic resonance, as there is no phase acquired on
resonance. Measurement of a phase imbalance at the interferometer
output cannot determine which qubit contributed the phase-shift,
resulting in entanglement between them.

\begin{figure}
    \includegraphics[height=65mm]{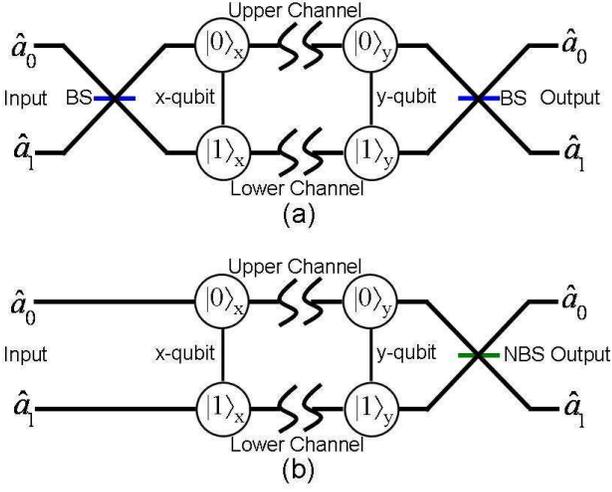}
    \caption{Schematic setup of entanglement generation with optical
    interferometers. Figure (a) shows the setup with the MZ interferometer
    which is consisted of two linear $50/50$ beamsplitters (BS). Figure
    (b) shows the setup with the NOON-state interferometer
    consisted of only one nonlinear beamsplitter (NBS).
 \label{fig1}}
 \end{figure}

We consider atomic qubits based on two degenerate hyperfine states,
arbitrarily labeled as $|0\rangle$ and $|1\rangle$. For a general
consideration, our goal is to entangle two uncorrelated qubits, labeled $x$ and $y$,
which are initially in states of $|\psi_x\rangle$, $|\psi_y\rangle$,
where
\begin{equation}
\label{initial}
    |\psi_\mu\rangle=\chi^\mu_0 |0\rangle_\mu+\chi^\mu_1 |1\rangle_\mu,
\end{equation}
and $\mu\in\{x,y\}$. The qubits are placed inside an optical
interferometer with the setup depicted in Fig.\ref{fig1}, where the
states $|0\rangle_x$ and $|0\rangle_y$ interact with photons in the
upper arm of the interferometer, while $|1\rangle_x$ and
$|1\rangle_y$ interact with the lower. Such interaction is
represented by the qubit-photon interaction propagator,
\begin{equation}
    \hat{U}_\mu=\exp[-i\theta(\ha_0^\dag\ha_0\hc_{\mu0}^\dag\hc_{\mu0}+\ha_1^\dag\ha_1\hc^\dag_{\mu
    1}\hc_{\mu 1})],
\end{equation}
 where  $\hat{c}_{\mu m}$ is the annihilation
operator for an atom at location $\mu\in\{x,y\}$ in internal state
$m\in\{0,1\}$. This interaction operator is valid in the
far-off-resonance regime, where the electronically excited state can
be adiabatically eliminated. The interaction is governed by the
phase-shift
\begin{equation}
\theta=\frac{|d{\cal E}(\omega)|^2\tau}{\hbar^2\Delta},
\end{equation}
where $\tau$ is the atom-photon interaction time, $\Delta$ is the
detuning between the laser and atomic resonance frequencies,  $d$ is
the electric dipole moment and ${\cal
E}(\omega)=\sqrt{\hbar\omega/(2\varepsilon_0V)}$ is the `electric
field per photon' for laser frequency $\omega$ and mode-volume $V$.
Introducing the spontaneous emission rate
$\Gamma=d^2\omega^3/(3\pi\varepsilon_0\hbar c^3)$, taking the photon
mode as having length $L$ and width $W$ (at the location of the
atom), and taking the interaction time as $\tau=L/c$, we arrive at
the single-atom phase-shift
\begin{equation}
\label{theta}
    \theta=\frac{3}{8\pi}\left[\frac{\lambda}{W}\right]^2 \frac{\Gamma}{\Delta},
\end{equation}
where $\lambda$ is the laser wavelength. This is the phase-shift
acquired by an off-resonant photon forward-scattered by a single
atom, and is independent of the pulse length.

The interferometer output is then determined by the phase-shift
acquired via the atom-photon interaction. Introducing the qubit-pair
basis $|ij\rangle\equiv |i\rangle_x\otimes |j\rangle_y$ with
$i,j=0,1$, the states $|01\rangle$ and $|10\rangle$ both correspond
to a balanced interferometer with zero net phase-shift, and thus
constitute a `balanced' qubit-pair subspace. In contrast, the states
$|00\rangle$ and $|11\rangle$ have equal and opposite non-zero
phase-shifts, and thus constitute an `imbalanced' subspace.
Measuring the photon number distribution at the interferometer
output distinguishes between zero and nonzero magnitudes of the
phase-shifts, and thus collapses the qubits onto the balanced or
imbalanced subspaces, based on which entanglement between the two is
established.

\section{MZ interferometer}
\label{cohrt}

The basic set-up for entanglement generation using MZ interferometer
is shown in Fig.\ref{fig1} (a). The MZ interferometer consists of
two 50/50 linear beamsplitters. The input light field is bifurcated
at the first beamsplitter, guided to interact sequently with the
qubits, and then recombined at the second beamsplitter. Passage of
photons through the MZ interferometer can be described by the
propagator,
\begin{equation}
    \hat{U}=\hat{U}_{BS}\hat{U}_y\hat{U}_x\hat{U}_{BS},
\end{equation}
where $\hat{U}_{BS}$ is the 50/50 beamsplitter propagator,
\begin{equation}
    \hat{U}_{BS}=\exp[-i(\ha_0^\dag\ha_1+\ha_1^\dag\ha_0)\pi/4].
\end{equation}
Without specifying the input light field, the initial states of the
system can be written in a general form
\begin{equation}
\label{input}
    |\Psi_i\rangle=\Phi(\ha^\dagger_0,\ha^\dagger_1)
    |0\rangle\otimes |\psi_x\rangle\otimes|\psi_y\rangle,
\end{equation}
where $|0\rangle$ is electromagnetic vacuum state and
$\Phi(\ha^\dagger_0,\ha^\dagger_1)$ defines the light field. The
state of the system at the interferometer output is then given by
\begin{eqnarray}
    |\Psi_f\rangle &=& \hat{U}|\Psi_i\rangle \\
    &=& \Phi(\hat{U}\ha^\dag_0\hat{U}^\dag,\hat{U}\ha^\dag_1\hat{U}^\dag)|0\rangle
    \otimes |\psi_x\rangle\otimes|\psi_y\rangle. \nonumber
\end{eqnarray}
Introducing dual-qubit spin operator
\begin{equation}
    \sigma_z=\frac{1}{2}\sum_{\mu=S, T} (\hc^{\dagger}_{\mu 0}\hc_{\mu
    0}-\hc^{\dagger}_{\mu 1}\hc_{\mu 1}),
\end{equation}
we find that
\begin{eqnarray}
    \hat{U}\ha^{\dagger}_0\hat{U}^\dagger=i e^{i\theta}[\sin (\theta
    \sigma_z) \ha^\dag_0+\cos(\theta\sigma_z) \ha^\dag_1], \\
    \hat{U}\ha^{\dagger}_1\hat{U}^\dagger=i e^{i\theta}[\cos (\theta
    \sigma_z) \ha^\dag_0-\sin(\theta\sigma_z) \ha^\dag_1].
\end{eqnarray}
The final state can now be rewritten as
\begin{equation}
\label{output}
    |\Psi_f\rangle=\sum_{i,j=0,1}
    \chi^x_{i}\chi^y_j |\Phi(\theta_{ij})\rangle\otimes |ij\rangle
\end{equation}
where $|\Phi(\theta_{ij})\rangle$ is the output light field in the
presence of qubits-dependent interferometer phase
$\theta_{ij}=\theta\times(1-i-j)$. It is now evident that the
interferometer output is determined by the joint states of the
qubits. The states $|01\rangle$ and $|10\rangle$ result in zero
phase-shifts with $\theta_{01}=\theta_{10}=0$, while $|00\rangle$
and $|11\rangle$ result in equation and opposite phases with
$\theta_{00}=-\theta_{11}=\theta$. If the interferometer is
incapable of distinguishing positive and negative phases, a
measurement of the output light field will therefore collapse the
qubits onto either balanced or imbalanced subspaces, and in this way
generate entanglement between them.

We note that our MZ interferometer scheme is closely related to the
Faraday-rotation scheme of Duan {\it et al}
\cite{DuaCirZol00}, which effectively replaces the first beamsplitter with a linear-polarized initial coherent state.
In fact, for the special case of a circularly polarized coherent state
at one input port and vacuum at the other, the present MZ
interferometry scheme maps directly to the Faraday-rotation
scheme. As we will show next, an interferometer of this class is limited to
shot-noise sensitivity, and will thus not work when the ensembles are replaced by single atoms without the introduction of extremely high-finesse optical resonators. Viewing the Faraday-rotation scheme instead as a MZ interferometer clearly highlights the possibility to incorporate non-classical input states to achieve sub-shot-noise sensitivity, which is the focus of the present manuscript.

\subsection{Coherent state input}
For the coherent-state input, the upper channel, described by
creation operator $\ha^\dag_0$, is initially in a coherent state,
while the lower channel $\ha^\dag_1$ is in the vacuum state. A
detector is used to count the photons coming from the upper output
channel, while output in the lower channel is unmeasured. A null
result, meaning zero photons detected, results in the qubits
collapsing onto the balanced subspace,
\begin{equation}
\label{balance-CH}
    \chi^x_0\chi^y_1 |01\rangle+\chi^x_1\chi^y_0
    |10\rangle+|\varepsilon\rangle,
\end{equation}
where $|\varepsilon\rangle$ is the intrinsic state error due to the
possibility of a false null result. This error, which adds
imbalanced states to the desired balanced subspace, sets the upper
limit of the obtainable teleportation fidelity. If $n\neq 0$ photons
are detected, the qubits will collapse onto imbalanced subspace
\begin{equation}
    \chi^x_0\chi^y_0 |00\rangle+\chi^x_1\chi^y_1 (-1)^n
    |11\rangle,
\end{equation}
without intrinsic error. We note that the possibility of a dark
count {\it will} introduce an analogous error, but this error rate
is governed by technical aspects of the photo-detector, and is
presumably not an intrinsic quantum error.

To derive these results for coherent input state, the initial state
of the complete system is given by equation (\ref{input}), with
\begin{eqnarray}
\label{coherentinput}
    \Phi_i(\ha^\dag_0,\ha^\dag_1)= e^{-\alpha\ha^{\dagger}_0+\alpha^\ast
    \ha_0}
\end{eqnarray}
Following equation (\ref{output}), the state of the system at the
interferometer output is obtained as
\begin{equation}
\label{output}
    |\Psi_f\rangle = \sum_{i,j=0,1}
    \chi^x_{i}\chi^y_j|ij\rangle\otimes |\bar{\alpha}\sin\theta_{ij}\rangle_0\otimes
    |\bar{\alpha}\cos\theta_{ij}\rangle_1
\end{equation}
where $\bar{\alpha}=-i\alpha e^{i\theta}$ and the states
$|\alpha\rangle_{0,1}$ indicate optical coherent states for the
upper and lower interferometer outputs, respectively. Expanding the
upper channel onto photon number-eigenstates and making the
small-angle approximation gives
\begin{equation}
\label{small-angle}
    |\Psi_f\rangle=\sum_{n=0}^\infty|n\rangle_0\otimes
    |\bar{\alpha}\rangle_1\otimes|\phi_n\rangle_{xy},
\end{equation}
where $|n\rangle_0$ indicates a state with $n$ photons in the upper
output, and
\begin{eqnarray}
\label{coherentoutput}
    |\phi_0\rangle_{xy}&=&\chi^x_0\chi^y_1|01\rangle+\chi^x_1\chi^y_0|10\rangle+|\varepsilon\rangle\\
    |\phi_{n\neq 0}\rangle_{xy}&=&f_n\left[\chi^x_0\chi^y_0|00\rangle + (-1)^n\chi^x_1\chi^y_1|11\rangle\right],
\end{eqnarray}
where
\begin{equation}
    |\varepsilon\rangle=e^{-|\alpha|^2\theta^2/2}
    (\chi^x_0\chi^y_0|00\rangle+\chi^x_1\chi^y_1|11\rangle)
\end{equation}
and
\begin{equation}
    f_n=\frac{\bar{\alpha}^n}{\sqrt{n!}} e^{-|\alpha|^2\theta^2/2}.
\end{equation}

The photon number in the upper channel is then measured with
single-photon resolution, while the output from the lower channel is
left unmeasured. From equation (\ref{small-angle}), the probability
of detecting $n$ photons $P(n)$ is given by
\begin{equation}
 P(n)=\Lambda\delta_{n,0}+ (1-\Lambda)e^{-N
\theta^2} \frac{N\theta^2)^{n}}{n!}
\end{equation}
where $\Lambda=|\chi^x_0\chi^y_1|^2+|\chi^x_1\chi^y_0|^2$ is the
weight of balanced-space states in the initial qubits' state. The
probability of detecting zero photons is thus
\begin{equation}
    P(0)=\Lambda(1-\varepsilon)+\varepsilon,
\end{equation}
where $\varepsilon=e^{-N \theta^2}$ indicates the probability of a
false null result. On detecting the null result, the qubits' state
will collapse onto
\begin{eqnarray}
& &|\Psi_B\rangle=\frac{1}{\sqrt{\Lambda(1-\varepsilon)+\varepsilon}}\times \\
&
&\left[\chi^x_0\chi^y_1|01\rangle+\chi^x_1\chi^y_0|10\rangle+\sqrt{\varepsilon}
(\chi^x_0\chi^y_0|00\rangle+\chi^x_1\chi^y_1|11\rangle)\right]
\nonumber
\end{eqnarray}
The fidelity upon this null result $f_{nul}$, which measures the
weight of balanced states in $|\Psi_B\rangle$, is thus
\begin{eqnarray}
f_{nul} &=& \frac{\Lambda}{\Lambda+(1-\Lambda)\varepsilon},
%& \approx& 1-\varepsilon\frac{1-\Lambda}{\Lambda},
\end{eqnarray}
which is non-unity due to the non-zero probability of a false null result.
The condition for faithful teleportation is then $\varepsilon\ll 1$,
or $N\theta^2\gg 1$, characteristic of a standard-quantum-limit
interferometer.

The remaining time, a photon-number $n\neq 0$ is detected, with the
qubit-state collapsing onto the imbalanced space with unit fidelity,
\begin{equation}
\label{unbl1}
    |\Psi_U\rangle=\frac{1}{\sqrt{1-\Lambda}}\left ( \chi^x_0\chi^y_0|00\rangle +
    (-1)^n\chi^x_1\chi^y_1|11\rangle \right ).
\end{equation}
The $(-1)^{n}$ term comes from the phase difference between
number-states for the coherent states $|\alpha\rangle$ and
$|-\alpha\rangle$, i.e. while measuring photon number can not
distinguish the states $|00\rangle$ and $|11\rangle$, it can
introduce relative phase between them. If the photon number is
definitely non-zero, yet not measured exactly, then tracing over the
photon number creates a statistical mixture of  $|00\rangle$ and
$|11\rangle$. In this case, the protocol would create an entangled
state with non-unity success probability $\Lambda$, but success would be heralded by
the verification of zero photons in the upper output. Most likely,
the initial state $\chi_m^\mu=1/\sqrt{2}$ would be prepared so that
$\Lambda=50\%$. For entanglement on-demand, however, it is necessary
to determine the photon number exactly. This difficulty is somewhat
mitigated by the fact that the average photon number is
$\bar{n}_0=-\ln \varepsilon$, i.e. only $5$ photons must be counted
for $\varepsilon=.01$ and $7$ for $\varepsilon=0.001$.

Leaving the lower output unmeasured means that computing the output
state requires tracing over the lower mode. In the proceeding
derivation we have taken this trace to be unity. In reality, it is
less than unity due to the non-orthogonality of the balanced and
imbalanced lower output states, governed by the overlap
\begin{eqnarray}
    |_1\langle
    \bar{\alpha}\cos\theta_{ii}|\bar\alpha\cos\theta_{ij}\rangle_{1}|^2
    &\approx& 1-(1-\delta_{ij})\theta^4 N/8 \nonumber\\
    &=& 1-O(1/N).
\end{eqnarray}
Here, $N=|\alpha|^2$ is the mean input photon number and the last
equality is because our scheme requires $N\theta^2\gg 1$. The
resulting error is then $\sim 1/N$, which can be neglected for large
$N$. This result validates the small-angle approximation made for
the final state as in (\ref{small-angle}), where the lower-channel
light field is assumed $\theta$-independent and factorized from the
remaining system.

The overall fidelity due to state error in this interferometrical
entanglement generation is obtained by averaging over the null- and
not-null- results, giving
\begin{eqnarray}
f_{avg}&=&P(0)\times f_{nul}+\sum P(n\neq 0) \times 1 \nonumber\\
  &=& 1-(1-\Lambda)\varepsilon
\end{eqnarray}
Since $\Lambda\ge 0$, it is always $f_{avg}\ge 1-\varepsilon$,
regardless of the quantum states of the two qubits.

Aside from the technical challenge of single-photon counting, the
fundamental quantum-mechanical barrier to successful teleportation
lies in finding a balance between phase-shift detection and
spontaneous-emission avoidance, as a single spontaneously scattered
photon can destroy the coherence of a qubit. The spontaneous
emission probability for a single qubit is $\theta N\Gamma/\Delta$,
which becomes negligible when $\theta N \Gamma/\Delta\ll 1$. This
condition must be satisfied without violating the
shot-noise-sensitivity condition $N\theta^2\gg 1$. From equation
(\ref{theta}) it follows that compatibility  requires
$16(W/\lambda)^2\ll 1$, which clearly violates the standard optical
diffraction limit. That such a scheme can therefore not work is in agreement
with common understanding \cite{EnkKim01}.

\subsection{Coherent state input with cavity feedback}

To overcome the effects of spontaneous emission, we can place the
two qubits in seperate high-finesse optical cavities, with
mechanical Q-switching employed to restrict the photon to $M$ passes
through each qubit. This will increase the phase-shift $\theta$ and
the spontaneous emission probability $P_{sp}$ by a factor of $M$.
This relaxes the compatibility condition to $8(W/\lambda)^2\ll M$,
which can be satisfied without sub-wavelength focussing.

The failure probabilities due to interferometry sensitivity and
spontaneous emission are then $\varepsilon=e^{-NM\theta^2}$ and
\begin{equation}
    P_{SP}=2NM\theta\Gamma/\Delta,
\end{equation}
respectively. Setting $P_{sp}=\varepsilon=.01$, corresponding to a
fidelity of $.99$, and taking $W/\lambda=3$ gives
\begin{eqnarray}
    M &=& -144\log{\varepsilon}/\varepsilon \\
      &=& 6.6\times 10^5, \nonumber
\end{eqnarray}
which is large but not necessarily outside the range of current
experimental techniques. For these parameters, the mean number of
photons in the upper output is $\bar{n}_0=4$, and the input photon
number is restricted only by the condition
$N\left(\Gamma/\Delta\right)^2=144\, \varepsilon/M= 4.4\times
10^{-6}$, together with the off-resonant condition
$\Delta\gg\Gamma$.

A main difficulty in long-distance quantum communication is photon
loss during qubit-to-qubit transmission, where the loss probability
increase exponentially with the transport distance. In schemes based
on cavity-QED \cite{BosKniPle99,ZheGuo06,DuaKim03,DiMutScu05},
atomic qubits' states are encoded in the internal (polarization)
states of photons, and thus a lost photon will immediately reveal
the atomic states and destroy the qubits via decoherence. In
contrast, during an interferometrical communication, the qubits'
state information is encoded in a form of relative phase-shifts of
photons propagating in the upper and lower arms. Such a shift is not
a measurable quantity until the two channels are recombined at a
second beamsplitter. Thus the lost photon cannot reveal the state of the qubit, and one might suspect that the qubit coherence would be preserved. On the
other hand, due to the photon-atom interaction, a lost photons will
introduce a small relative phase-shift to the qubits. The magnitude of the relative phase is
$\theta$, but the sign depends on which interferometer `arm' lost the photon.
Tracing over which arm thus
results in effective decoherence and thus
 a reduction in the fidelity of
entanglement.

To see this, we first consider one photon lost during propagating
between the first and the second qubits. This will alter the final
state into
\begin{equation}
    |\Psi'_f\rangle=\sqrt{\frac{2}{N}}\hat{U}_{BS}\hat{U}_y\ha_q
    \hat{U}_x \hat{U}_{BS} |\Psi_i\rangle,
\end{equation}
with $q=0,1$ corresponding to the loss in upper and lower arms,
respectively. The identity
\begin{equation}
\label{iden}
    \hat{U}_{BS} \hat{U}_{y} \ha_q \hat{U}^\dag_y
    \hat{U}^\dag_{BS}=e^{i\theta \hc^{\dagger}_{yq}\hc_{yq}}
    (\ha_q-i\ha_{1-q})/\sqrt{2},
\end{equation}
enables us to write,
\begin{eqnarray}
\label{lofi}
    |\Psi'_f\rangle &=& \sqrt{\frac{1}{N}} e^{i\theta \hc^{\dagger}_{yq}\hc_{yq}}
    (\ha_q-i\ha_{1-q})\hat{U}|\Psi_i\rangle \\
    &=& (-i)^q e^{i\theta \hc^{\dagger}_{yq}\hc_{yq}}|\Psi_f\rangle,
\end{eqnarray}
where in the last step we have used the fact that for the present
input state (\ref{coherentinput}), $\ha_q|\Psi_i\rangle=\alpha
\delta_{q,0}|\Psi_i\rangle$. It is now clear that the net effect of one lost photon is
equivalent to introducing a relative phase $\pm\theta$ to the qubit, where $\theta\ll 1$. 
In the case of random photon losses, such phase
disturbances will lead to the unknown drift of the qubit's state and
thus a reduction in the overall fidelity of the entanglement
generation. To estimate this fidelity reduction, we introduce the
lost  photon number distribution $f(k)$. Because each photon is lost
independently, $f(k)$ will exhibit a Poisson distribution, where for
a mean loss number $\bar{k}$, the variance is $\sqrt{\bar{k}}$. For
simplicity, we approximate $f(k)$ with a gaussian,
\begin{equation}
\label{lost}
    f(k)=\frac{e^{-(k-\bar{k})/2\bar{k}}}{\sqrt{2\bar{k}\pi}}.
\end{equation}
The system's density $\rho^{loss}$ after the loss is then a mixture
of
\begin{equation}
    \label{loss} \rho^{loss}=\sum_{k,k'} (-i)^{k'}f(k)
    f(k')(\hat{P}_{0})^k(\hat{P}_{1})^{k'} |\Psi_f\rangle\langle\Psi_f|
    (\hat{P}^\dagger_{1})^{ k'} (\hat{P}^\dagger_0)^{k},
\end{equation}
where we have introduced the $y$-qubit projector
\begin{eqnarray}
\label{proj}
    \hat{P}_{q} &=& e^{i\theta \hc^{\dagger}_{yq}\hc_{yq}} \nonumber\\
    &=& e^{i\theta}|q\rangle_y\langle q|+|1-q\rangle_y\langle
    1-q|.
\end{eqnarray}
Defining the reduced fidelity due to the photon loss,
\begin{equation}
    f_{loss}=tr\{\rho\rho^{loss}\},
\end{equation}
it is found
\begin{equation}
    \bar{k}=\frac{N\ln(2f_{loss}-1)}{\ln \epsilon}.
\end{equation}
Taking $f_{loss}=1-\epsilon=.99$ gives $\bar{k}=.004 N$, meaning
about one photon can be lost in every $250$ photons.

In conclusion, in this section we showed the MZ-interferometrical
generation of entanglement using coherent state is
quantum-mechanically allowed only with the aid of optical
resonators. We found that a fidelity of $.99$ can be achieved using
ring cavities which cycle photons for $6.6\times 10^5$ times, with
about $4$ photons needing to be measured accurately at one output.
Furthermore, we found that unlike most cavity-QED schemes, the
present approach can be tolerant of a small photon loss rate.

\begin{figure}
 \includegraphics[height=55mm]{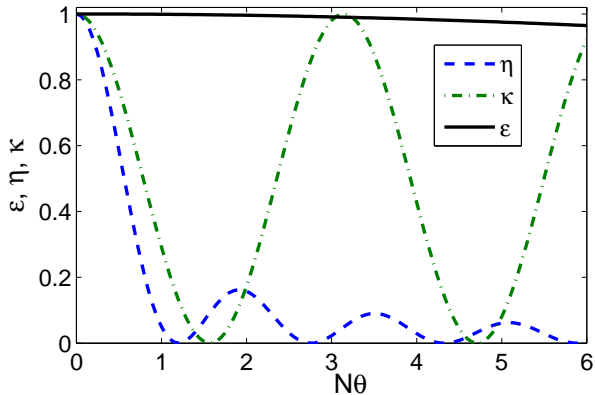}.
 \caption{(color online) An example of intrinsic error due to interferometer sensitivity.
  Errors for MZ interferometer with coherent $\varepsilon$ (solid), FT input $\eta$ (dashed),
and NOON-state interferometer $\kappa$ (dashed-dotted) are plot as
functions of $N\theta$ (with $N=10^3$), respectively. Note while
$\varepsilon$ is dependent on $N\theta^2$, both $\eta$ and $\kappa$
are dependent on $N\theta$.
 \label{fig2}}
 \end{figure}

\subsection{Twin-Fock state input } \label{twinfock} To
achieve a higher fidelity, and/or to eliminate the need for a
high-finesse resonator, we now consider using sub-shot-noise
interferometers to overcome the spontaneous emission to phase
sensitivity. In this section, we investigate the fundamental limits
when a twin-Fock (TF) photon input state is used to increase the
phase sensitivity of the MZ interferometer. The TF input set-up
differs in that the photon number-difference between the outputs
must be measured. In this case, a result of zero number-difference
constitutes a null result.

The input state is now $|N,N\rangle$, with the dual-Fock basis
defined as
\begin{equation}
    |k,l\rangle=(\ha^\dagger_0)^k(\ha^\dagger_1)^l|0\rangle/\sqrt{k!~l!}.
\end{equation}
The TF input state is then $|N\rangle_0|N\rangle_1$ with $0,1$
corresponding to upper and lower inputs as before. Following
equation (\ref{output}), the output state is now
\begin{equation}
\label{df}
    |\Psi_f\rangle = \sum_{ij}\chi^x_i\chi^y_j|ij\rangle\otimes\sum^{N}_{m=-N}
    \xi_m(\theta_{ij}) |N+m, N-m\rangle ,
\end{equation}
where
\begin{eqnarray}
    \xi_m(\theta_{ij}) =\sum^{\min\{N, N-m\}}_{l=\max\{0, -m\}} (-1)^{m+l} \
    \left(\begin{array}{c}m+l \\N \end{array}\right)
    \left(\begin{array}{c}l \\N \end{array}\right) \times \nonumber \\
    \frac{\sqrt{(N+m)! (N-m)!}}{N!}(\sin\theta_{ij})^{m+2l}
    (\cos\theta_{ij})^{2N-m-2l}. \nonumber
\end{eqnarray}
The desired two-qubit entangled state is then created by measuring
the photon number difference between the upper and lower outputs. It
is seen from Eq. (\ref{df}) that the probability of detecting a
difference of $2m$ is given by
\begin{equation}
   P(2m)=\Lambda\delta_{n,0}+ (1-\Lambda)\xi^2_m(\theta),
\end{equation}
where again $\Lambda=|\xi^x_0\chi^y_1|^2+|\chi^x_1\chi^y_0|^2$. The
probability to detect zero photon number difference (or a null
result) is thus
\begin{equation}
    P(0)=\Lambda(1-\eta)+\eta,
\end{equation}
where $\eta=\xi^2_0(\theta)$ is the probability of a false null
result. On detecting the null result, the qubit state will collapse
onto
\begin{eqnarray}
\label{balance-TF}
& &|\Psi_B\rangle=\frac{1}{\sqrt{\Lambda(1-\eta)+\eta}}\times \\
&
&\left[\chi^x_0\chi^y_1|01\rangle+\chi^x_1\chi^y_0|10\rangle+\sqrt{\eta}
(\chi^x_0\chi^y_0|00\rangle+\chi^x_1\chi^y_1|11\rangle)\right],
\nonumber
\end{eqnarray}
with the corresponding fidelity
\begin{eqnarray}
f_{nul} &=& \frac{\Lambda}{\Lambda+(1-\Lambda)\eta}.
%& \approx& 1-\varepsilon\frac{1-\Lambda}{\Lambda},
\end{eqnarray}
The remaining time, a photon number difference $m\neq 0$ is
detected, with the qubits collapsed to
\begin{equation}
\label{unbalance}
 |\Psi_U\rangle=\chi^x_0\chi^y_0|00\rangle +
(-1)^m\chi^x_1\chi^y_1|11\rangle.
\end{equation}
Here, similar to the coherent state, the exact photon number
difference must be measured in order to successfully disentangle the
qubits. The overall fidelity in this entanglement generation is then
\begin{eqnarray}
    f_{avg}&=&1-(1-\Lambda)\epsilon \nonumber\\
            &\ge& 1-\epsilon.
\end{eqnarray}

The TF input thus yields results similar to the coherent
state-input, but with the intrinsic error due to interferometer
sensitivity given by $\eta$ instead of $\varepsilon$.  A comparison
plot of $\eta$ and $\varepsilon$ is shown in Fig. \ref{fig2}, where
it is seen that $\eta$ decreases with $N$ much faster than
$\varepsilon$. In fact, for $N\theta < 1$, $\eta\approx e^{-N^2
\theta^2}$, which is characteristic of a Heisenberg-limited phase
sensitivity.  This means that significantly fewer photons are
required to obtain equal fidelity, with a corresponding reduction in
spontaneous emission. In Fig. \ref{fig2}, we see that the false-null
probability $\eta$ is exactly zero for a periodic set of values of
$N\theta$. The first such zero occurs at $N\theta=1.196\equiv x_1$.
Thus if one can precisely control $N\theta$, {\it it is possible to
achieve teleportation without intrinsic error due to false-null
results}. In this case, the success of teleportation is governed
only by spontaneous emission probability $P_{sp}=2 N \theta
\Gamma/\Delta=2 x_1\Gamma/\Delta$. The condition $N\theta=x_1$,
together with (\ref{theta}), means that $\Gamma/\Delta=(8
x_1/N)\left(W/\lambda\right)^2$, so that
\begin{equation}
\label{spt} P_{sp}=(16 x_1^2/N)\left(W/\lambda\right)^2.
\end{equation}
For the case of a tightly focussed beam, we can take
$W/\lambda\approx 3$ this gives $P_{sp}=206/N$. The theoretical
limit to fidelity therefore scales as $\sim1-200/N$, thus a fidelity
of $f=.99$ would require $N=2\times 10^4$ (or a total of $4\times
10^4$ photons), while a fidelity of $f=.999$ could be achieved with
$N=2\times 10^5$. An extremely high fidelity of $f=.999999$ would
therefore require $N=2\times10^8$. The addition of a Q-switched
cavity with $M$ cycles replaces $N$ with the effective photon number
$MN$, resulting in the spontaneous emission probability results in
$P_{sp}=206/(NM$), which for $M=2\times 10^4$, would reduce the
photon numbers to $N=1$ for $f=.99$, $N=10$ for $f=.999$, and
$N=10^4$ for $f=.999999$.

The exactly elimination of false-null-induced reduction in fidelity
requires the precise control of single-particle phase shift
$\theta$, as well as the particle number $N$. Imprecise controls of
either will lead to $\eta\neq 0$, and thus a reduction in overall
fidelity. To estimate this effect, we let $N\theta=x_1+\delta$, with
$\delta$ resulted from the displacement of $\theta$ and/or $N$.
Expanding $\eta(x_1+\delta)$ near $\eta(x_1)=0$ gives
\begin{equation}
    \eta(x_1+\delta)\approx 1.3\ \delta^2,
\end{equation}
For a fidelity of $f=1-200/N$ (with $\eta=200/N$), it requires
$\delta<12.4/\sqrt{N}$. This then requires  $\theta_1-9.5\
\theta^{1.5}_1<\theta<\theta_1+9.5\ \theta^{1.5}_1$, where
$\theta_1=x_1/N$ is the desired per-atom phase shift. This allows a
relatively flexible control of $\theta$.

\begin{figure}
 \includegraphics[height=55mm]{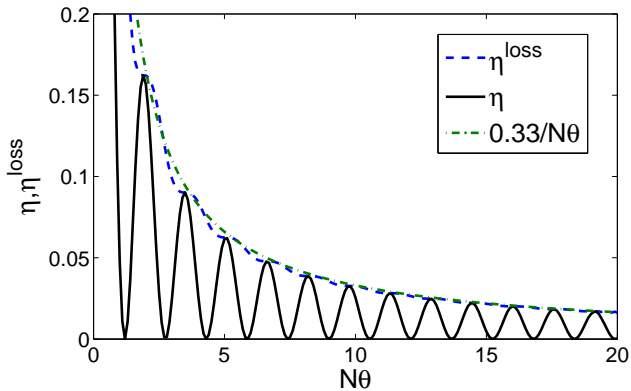}.
 \caption{Comparison of $\eta^{loss}$ and $\eta$. Note that both
 depend only on the product $N\theta$.
 \label{fig3}}
 \end{figure}

Lastly, we note that for the TF input and the present parameter
choice of $N\theta=x_1$, a single photon loss will immediately
reduce the fidelity, with a worst-case result of $f=0.73$ and thus disrupt the on-demand
entanglement generation scheme. This is because a lost photon will
lead to a rapid degradation of the interferometer sensitivity. To see this,
for the TF input state, one photon lost from the $q$-th path during
qubit-to-qubit propagation will result in the final state
\begin{equation}
    |\Psi'_f\rangle=\sqrt{\frac{1}{N}}\hat{U}_{BS}\hat{U}_y\ha_q
    \hat{U}_x \hat{U}_{BS} |\Psi_i\rangle,
\end{equation}
Use the identity (\ref{iden}), we find
\begin{eqnarray}
    |\Psi'_f\rangle &=& \sqrt{\frac{1}{2N}}\left[e^{i\theta
    \hc^{\dagger}_{yq}\hc_{yq}} (\ha_q-i\ha_{1-q})\right] \hat{U} |\Psi_i\rangle, \\
    &=& \sqrt{\frac{1}{2N}} \sum_{ij}\chi^x_i\chi^y_j \hat{P}_q|ij\rangle\otimes\sum^{N}_{m=-N}
    \xi_m(\theta_{ij}) \times \nonumber\\
     & & [(-i)^q\sqrt{N+m}|N+m-1, N-m\rangle+\nonumber\\
     & & (-i)^{1-q}\sqrt{N-m}|N+m, N-m-1\rangle], \nonumber
\end{eqnarray}
with the projector $\hat{P}_q$ defined in (\ref{proj}). A lost
photon will therefore result in odd number differences of photons
measured in the two output ports. Since without photon loss, a TF
state will always result in even number differences, it is in this
way possible to determine the loss of a single photon (while without
knowing which path it is lost from). Seemingly, this makes it
possible to detect the phase imbalance if we accordingly redefine a
null result as the measured photon number difference being $1$. The
false-null rate as given by
\begin{equation}
    \eta^{loss}=|\xi_0(\theta)+i\xi_{1}(\theta)\sqrt{1+1/N}|^2,
\end{equation}
is, however, no long a small quantity. A comparison of $\eta^{loss}$
and $\eta$ is plotted in Fig. \ref{fig3}, where it is shown
$\eta^{loss}$ behaves as the envelope of $\eta$ without the
zero-value points. In particular, with the present choice of
$N\theta=x_1$, it is found $\eta^{loss}\approx 0.27$, in contrast to
the corresponding rate $\eta=0$ without the loss. Depending on the
qubits' states,  a photon loss will thus immediately degrade the
fidelity to $f\ge 0.73$.

On the other hand, $\eta^{loss}$ is yet much smaller than the
corresponding false-null rate $\varepsilon$ for the coherent state.
Hence, if we presume one photon will be lost and set the value of
$N\theta$ accordingly, we may still generate entanglement without
the cavity enhancement. In fact, as shown in Fig. \ref{fig3}, a
least-square fit finds
\begin{equation}
    \eta^{loss}=\frac{0.33}{N\theta}.
\end{equation}
Letting $\eta^{loss}=P_{sp}$ and using equation (\ref{theta}) gives
\begin{equation}
    P_{sp}=\frac{2.6}{N^{1/3}},
\end{equation}
for $W/\lambda=3$. The limit to fidelity thus scales as
$1-2.6/N^{1/3}$, and a fidelity of $f=.99$ will require $N=1.8\times
10^7$, compared to $N=2\times 10^4$ without photon loss.
Entanglement can in this sense still be generated without the need
of cavities, while the single-photon loss can be compensated by
using more photons. Lastly, we note that $f\sim 1-2.6/N^{1/3}$ is
also the lower limit on the fidelity achievable when the phase shift
can not be tuned such that $\eta(N\theta)=0$. This is simply because
$\eta^{loss}$ is the envelop of $\eta$, and for any $N$ and
$\theta$, $\eta\le \eta^{loss}$.

In conclusion, in this section we have shown that for a MZ
interferometer with the TF input, atom-atom entanglement can be
generated with much higher fidelity, and the need for high-finesse
optical resonators can in principle be eliminated. Particularly, we
found a fidelity of $.99$ is quantum-mechanically allowed with
$20,000$ photons, or more intriguingly with only $2$ photons, provided
ring cavities which cycle photons
$2\times 10^4$ times are additionally incorporated. The 2-photon TF state could be generated with
a pair of single-photon-on-demand sources (one for each input) and a
precise photon detector to measure the two-photon output state,
technologies that are rapidly advancing at present. Finally, we have
shown the present scheme is relatively insensitive to deviations in
the per-atom phase shift, yet is highly sensitive to loss of a
single photon. This is somewhat mitigated by the fact that for the
2-photon state, the loss of a photon could be readily detected, so
that success is heralded by the detection of both photons.

\section{NOON-state interferometer}
\label{noon}

In above sections, we have discussed generating entanglement between
atomic qubits using an optical MZ interferometer with coherent and
TF input states. While both are shown to be able to achieve a
close-to-unit fidelity in the presence of intrinsic quantum errors,
they require precise measurement of output light field at the
single-photon level. In this section, we show this requirement can
be overcome by using a non-MZ interferometer based on NOON states
and nonlinear beamsplitters \cite{KimDunBur05}.

A NOON state is a Shr\"{o}dinger cat state that corresponds to an
equally-weighted superposition of all-upper-channel and
all-lower-channel states
\cite{BolItaWin96,ZouKimLee01,GerCam01,LeeKokCer02,MitLunSte04}
\begin{equation}
    |\mathrm{NOON}\rangle=\frac{1}{\sqrt{2}}(|N,0\rangle+e^{i\phi}|0,N\rangle),
\end{equation}
where $\phi$ is the relative phase which we take for zero for
simplicity. The nonlinear beamsplitter can either be a four-wave
mixer \cite{YurMcCKla85,Ger00} or a quantum circuit constructed from
C-Not gates \cite{NieChu00}. Without further explaining its
operational mechanism or examining the practical feasibility, for
present we simply treat the action of such beamsplitters with a
projecting operator $\hat{U}_{NBS}$ of the general form
\begin{eqnarray}
    \hat{U}_{NBS}&=&\frac{1}{\sqrt{2}} \sum_{s}(e^{i\phi}|s,0\rangle+
    |s,0\rangle)\langle 0,s|\nonumber\\
    & &+e^{i\varphi} (e^{i\phi}|s,0\rangle-|0,s\rangle)\langle 0,s|,
\nonumber
\end{eqnarray}
where for simplicity we let the relative phases $\phi=\varphi=0$.
Note for the four-wave mixer, this projector is valid only for
even-number $s$. Using this nonlinear beamsplitter, a NOON state can
be generated from a single Fock state $|N,0\rangle$, i.e., by
injecting $N$ photons in its upper input channel.

The set-up of the NOON-state interferometer differs from a MZ
interferometer in that now the first beamsplitter is dropped (or
more precisely, it is formally replaced by the assumption of a NOON
input state) while the second one is replaced by the nonlinear
beamsplitter, as shown in Fig.\ref{fig1} (b). For measurement, a
photo detector is placed in the lower (or equivalently, the upper)
output port to detect the presence of outcoming photons, while
without counting them. Because the output light field is consisted
of $|N,0\rangle$ and $|0,N\rangle$ states, corresponding to all
photons coming out from upper or lower port, a photon detector with
a resolution of $<N/2$ would be sufficient to distinguish them. This
exhibit an essentially improvement from the previous MZ
interferometry schemes, where the detector resolution must be less
than one. A null result, meaning no photon is detected at the lower
channel, will collapse the qubits onto
\begin{equation}
    \chi^x_0\chi^y_1|01\rangle+\chi^x_1\chi^y_0|10\rangle+|\kappa\rangle,
\end{equation}
with $|\kappa\rangle$ the intrinsic state error due to a false-null
result. In contrast, if the detector is trigged, the qubits will
collapse into the imbalanced subspace
\begin{equation}
    \chi^x_0\chi^y_0|00\rangle+\chi^x_1\chi^y_1|11\rangle.
\end{equation}
Here, we emphasize that this state is independent on the exact
number difference $n$ between the upper and lower outputs. This is
essentially different from the corresponding ones in (\ref{unbl1})
with the coherent state and in (\ref{unbalance}) with the TF state,
both of which are dependent on $n$.

To derive these results, we follow the previous approach and find
the final state of system as
\begin{eqnarray}
& & |\Psi_f\rangle = \hat{U}_{NBS}\hat{U}_y\hat{U}_x \left[
\frac{1}{\sqrt{2}}(|N,0\rangle+|0,N\rangle)\otimes
|\psi_x\rangle \otimes |\psi_y\rangle\right] \nonumber \\
& &= \sum_{i,j=0,1} \chi^{x}_i\chi^{y}_j|ij\rangle \otimes [\cos
N\theta_{ij}|N,0\rangle-i\sin N\theta_{ij}]0,N\rangle] ,\nonumber
\end{eqnarray}
where we have dropped an irrelevant global-phase term in the last
step. The probability of a null result is thus
\begin{equation}
P_{null}=\Lambda(1-\kappa)+\kappa,
\end{equation}
with $\kappa=\cos^2 N\theta$ is the intrinsic error rate. On
detecting this null result, the qubits will collapse onto
\begin{eqnarray}
\label{blan-noon}
& &|\Psi_B\rangle=\frac{1}{\sqrt{\Lambda(1-\kappa)+\kappa}}\times \\
& &
\left[\chi^x_0\chi^y_1|01\rangle+\chi^x_1\chi^y_0|10\rangle+\sqrt{\kappa}
(\chi^x_0\chi^y_0|00\rangle+\chi^x_1\chi^y_1|11\rangle)\right]
\nonumber
\end{eqnarray}
where $\kappa=\cos^2 N\theta$ is the false null rate. The
corresponding fidelity is then
\begin{equation}
f_{nul}=\frac{\Lambda}{\Lambda+(1-\Lambda)\kappa}.
\end{equation}
The remaining time, a not-null result is detected, projecting the
qubits onto the imbalanced subspace,
\begin{equation}
\label{imb-noon}
    |\Psi_U\rangle=\frac{1}{\sqrt{1-\Lambda}}
    \left ( \chi^x_0\chi^y_0|00\rangle+\chi^x_1\chi^y_1|11\rangle \right ).
\end{equation}
The overall fidelity averaging over the null and not-null results is
given by
\begin{equation}
    f_{avg}=1-(1-\Lambda)\kappa.
\end{equation}

The fidelity in this entanglement generation is thus similar to the
MZ interferometry cases but with the intrinsic error rate given by
$\kappa$. A comparison of $\varepsilon, \eta, \kappa$ is shown in
Fig. \ref{fig2}, where it is shown that for $N\theta<1$, $\kappa$ is
close to $\eta$, exhibiting a Heisenberg-limited phase sensitivity.
Furthermore, $\kappa$ is exact zero at $N\theta=\pi/2$, compared to
at $1.196$ for $\eta$. This means the NOON-state interferometer can
achieve similar performances with the TF state. The fidelity after
taking into account the possibility of spontaneous emission
therefore scales as $f\approx 1-350/N$. A fidelity of $f=.99$ and
$f=.999$ will then require $N=3.5\times 10^4$ and $N=3.5\times 10^5$
photons, respectively. Since there is no requirement on exactly
counting output photons, $N$ can in principle be made large,
allowing an arbitrary close-to-unit fidelity, at least quantum
mechanically. Finally, similar to TF state, in order to suppress
false-null rate, for a fidelity of $\approx 1-350/N$, $\theta$ must
be tuned within an interval of $(\theta_1-8.1
\theta^{1.5}_1,\theta_1+8.1 \theta^{1.5}_1)$ with $\theta_1=\pi/2N$.

The present entanglement generation using NOON states will be
completely disrupted by a single-photon loss. This is due to the fact that a
randomly lost photon will immediately collapse the NOON state to a
statistical mixture of all-upper-channel and all-lower-channel
states, whose reduced density is given by
\begin{equation}
    \rho^{loss}=\frac{1}{2}(|N-1,0\rangle\langle
    N-1,0|+|N-1,0\rangle\langle N-1,0|).
\end{equation}
This mixture state is apparently incapable of detecting phase
imbalances. This problem, however, might be overcome by using a
class of less-extreme cat-like states \cite{MicJakCir03,HuaMoo06}.
Such a states corresponds to a symmetric superposition of two
well-separated wavepackets in number-difference space. In the case
of small photon losses, instead of being completely destroyed, they
will decay into a mixture of smaller-sized cat-like states, which
are still suitable for the purpose of detecting phase imbalance.
Hence, asides from a reduction in fidelity due to phase
randomization, faithful entanglement might be generated despite of
photon loss. A further study of generating entanglement using
less-extreme cat-like states will be presented in future work.

Finally, we note that the present NOON-state interferometer relies
on highly nonclassical light source with definite photon number.
This requirement is, however, not a necessary. For example, our
scheme can be directly extended to use more 'classical' cat input
states that correspond to the superposition of coherent states,
\begin{equation}
    |\alpha\rangle_0 \otimes |0\rangle_1+|0\rangle_0 \otimes
    |\alpha\rangle_1.
\end{equation}
Such a state can be rewritten as,
\begin{equation}
    \sum_m f(m) (|m,0\rangle+|0,m\rangle).
\end{equation}
where $f(m)$ is the coefficient of the coherent state and can be
approximated by
\begin{equation}
f(m)=\frac{1}{\sqrt{\mathcal{N}}} e^{-(m-N)^2/4N}.
\end{equation}
where $\mathcal{N}$ is the normalization factor. A state yielding
this distribution but containing only even $m$s has been proposed to
generate probabilistically with a success probability of half, where
a single coherent light and a four-wave mixer are employed
\cite{Ger00}. With this input, the final state of the system becomes
\begin{eqnarray}
    |\Psi_f\rangle &=& \sum_{i,j=0,1} \chi^{x}_i\chi^{y}_j|ij\rangle
    \otimes \\
    & & \sum_m f(m)[\cos m\theta_{ij}|m,0\rangle-i\sin
    m\theta_{ij}]0,m\rangle],\nonumber
\end{eqnarray}
By choosing $N\theta=\pi/2$, similar to the single NOON state, upon
detecting photons from the lower port, the qubit will collapse to
imbalanced subspace (\ref{imb-noon}). Otherwise, if no photon is
detected from the lower port, the qubits will collapse to the state
(\ref{blan-noon}) but with the false null rate given by
\begin{eqnarray}
    \kappa' &=& \frac{1}{2\mathcal{N}} \sum_m e^{-(m-N)^2/2N} (1+\cos
    m\theta) \\
      &=& \frac{1}{2}(1-e^{-\theta^2N/2}) \nonumber
\end{eqnarray}
Since $N\theta=\pi/2$, we have $\kappa'\approx\pi^2/16N$, which is
negligible compare to the fidelity reduction ($\approx 350/N$) due
to the probability of spontaneous emission.

To conclude, in this section, we have shown that a NOON-state
interferometer can achieve similar performance with the TF state,
yet without the requirement for precisely measuring the output light
field. While this scheme is not tolerant of a single photon loss,
this problem might be overcome by using a class of less extreme cat
states. Also, besides using a single NOON input state with definite
photon number, we showed the present scheme can also use a class of
states with indefinite photons that correspond to the superposition
of NOON states.

\section{Examples of applications}
\label{app} The present interferometrical method of generating
entanglement can serve as a basic protocol in the quantum
information processing, based on which quantum computation and
communication can be realized with the aid of local qubit
operations. As an example, here we first show how it can be used to
teleport an arbitrary quantum state from one qubit to another. We
assume the $x$-qubit is the source qubit carrying an unknown
teleporting quantum state, and $y$-qubit is target qubit which the
state is transported to. The $x$-qubit is initially in the state
\begin{equation}
    |\psi_x\rangle=\chi^x_0|0\rangle_x+\chi^x_1|1\rangle_x,
\end{equation}
while the $y$-qubit is initially prepared as
\begin{equation}
    |\psi_y\rangle=\frac{1}{\sqrt{2}} (|0\rangle_y+|1\rangle_y).
\end{equation}
We first collapse the two qubits into entangled qubit-pair
interferometrically using our method. Once the qubit-pair is
generated, completing the teleportation requires that the qubits be
disentangled. This can be accomplished in the following manner.
Conditional upon a null result, a $\pi$-pulse is applied to the
source qubit, flipping $|0\rangle_x \leftrightarrow |1\rangle_x$.
When using a MZ interferometer with the coherent or TF state, in the
case of an odd measured $n$, an additional relative $\pi$ phase must
be applied to the state $|1\rangle_{x}$ (or $|1\rangle_{y}$). After
these steps, the qubits' state becomes
\begin{equation}
    \chi^x_0|00\rangle+\chi^x_1|11\rangle.
\end{equation}
A $\pi/2$-pulse is then applied to the source (or the target) qubit,
transforming the state into
\begin{equation}
    [\chi^x_0|00\rangle-i\chi^x_0|10\rangle-i\chi^x_1|01\rangle+\chi^x_1|11\rangle]/\sqrt{2}.
\end{equation}
This is followed by a state measurement of $x$-qubit. If it is
measured $|0\rangle_x$, the $y$-qubit will collapse to
\begin{equation}
    \chi^x_0 |0\rangle_y-i \chi^x_1 |1\rangle_y,
\end{equation}
after which a $\pi/2$ phase is imprinted onto $|1\rangle_y$.
Otherwise, it is measured in $|1\rangle_x$, and a $\pi/2$ phase is
imprinted onto $|0\rangle_y$. After these conditional operations,
the $y$-qubit will end up in the desired state
\begin{equation}
    \chi^x_0|0\rangle_y+\chi^x_1|1\rangle_y,
\end{equation}
which accomplishes the teleportation.

Besides the state teleportation between two qubits, our scheme can
be easily generalized to generate many-qubit entanglement
\cite{GreHorZei89} as well as realize  entanglement swapping
\cite{PanBouWei98,DuaLukCir01}. For example, the three-particle
Greenberger-Horne-Zeilinger (GHZ) state,
\begin{equation}
    (|000\rangle_{xyz}+|111\rangle_{xyz})/\sqrt{2},
\end{equation}
can be created by first preparing each qubit in the state
\begin{equation}
    |\psi_\mu\rangle=\frac{1}{\sqrt{2}}(|0\rangle_i+|1\rangle_i),
\end{equation}
with $i=x,y,z$. Then the two-qubit protocol is used to collapse $x$
and $y$ into the state
\begin{equation}
    (|00\rangle_{xy}+|11\rangle_{xy})/\sqrt{2}\otimes
    (|0\rangle_z+|1\rangle_z)/\rangle/\sqrt{2}.
\end{equation}
\\ If the same two-qubit procedure is applied to $B$ and $C$, the GHZ
state is obtained. This simple scheme can be extended in a
straightforward manner to producing an $N$-particle Shr\"{o}dinger
cat state.

To realize entanglement swapping, we take an initially entangled
qubit-pair,
\begin{equation}
    \frac{1}{\sqrt{|c_{00}|^2+|c_{11}|^2}}(c_{00}|00\rangle_{xy}+c_{11}|11\rangle_{xy}),
\end{equation}
and an uncorrelated third qubit
\begin{equation}
     |\psi_z\rangle=\frac{1}{\sqrt{2}}(|0\rangle_z+|1\rangle_z)
\end{equation}
and apply our protocol to qubits $y$ and $z$ to create a GHZ-like
state. Then, by disentangling $y$ in the same manner as described
for the source qubit in teleportation, we arrive at the desired
swapped state
\begin{equation}
        \frac{1}{\sqrt{|c_{00}|^2+|c_{11}|^2}}(c_{00}|00\rangle_{xz}+c_{11}|11\rangle_{xz}).
\end{equation}

\section{conclusion}
\label{conclusion} In conclusion, we have used the formalism of the
optical interferometer to treat the problem of creating entanglement
amongst single-atom qubits via a common photonic channel. We have
compared the results from a MZ-interferometer with a coherent input
state and high-finesse cavity enhancement, a MZ-interferometer with
TF input and those from a non-MZ interferometer based on the NOON
state and nonlinear beamsplitter. Our results suggest that
high-fidelity entanglement can in principle be generated via any of
the interferometrical approaches. Experimental feasible schemes
under current techniques are found by combining Hisenberg-limited
interferometer with photon resonators. In particular, we find that a
two-photon input state has a fundamental upper-limit to fidelity of
$.99$, and provides the advantage that failure due to photon losses
could be readily detected. Our interferometrical approaches of
generating entanglement is operated on-demand and is scalable, and
thus can serve as a universal protocol in quantum information
processing, based on which quantum computation and communication can
be realized with the aid of single-qubit operations.

This work is supported in part by Nation Science Foundation Grant
No. PHY0653373.


\begin{thebibliography}{99}

\bibitem{NieChu00} M. A. Nielsen and I. L. Chuang, \emph{Quantum Computation and
Quantum Information} (Cambridge University Press, Cambridge, United
Kingdom, 2000).

\bibitem{BenBraCre93} C. H. Bennett, G. Brassard, C. Crepeau, R. Jozsa, A. Peres, and W. K.  Wootters, Phys. Rev. Lett. {\bf 70}, 1895 (1993).

%\bibitem{CHB} C. H. Bennett and D P. DiVincenzo, Nature {\bf 404}, 247 (2000).

\bibitem{BouPanMat97} D. Bouwmeester, J.-W. Pan, K. Mattle, M. Eibl, H. Weinfurter, and A. Zeilinger, Nature {\bf 390}, 575 (1979);

%\bibitem{Fat04}  D. Fattal $et~ al.$, Phys. Rev. Lett. {\bf 92}, 037904 (2004).

\bibitem{MarRieTit03} I. Marcikic, H. de Riedmatten, W. Tittel, H. Zbinden, and N. Gisen, Nature {\bf 421} 509 (2003).

\bibitem{BosKniPle99}  S. Bose, P. L. Knight, M. B. Plenio, and V. Vedral, Phys. Rev. Lett. {\bf 83}, 5158 (1999).

\bibitem{ZheGuo06}  S.-B. Zheng and G.-C. Guo,  Phys. Rev. A {\bf 73}, 032329 (2006);

 \bibitem{DuaKim03} L.-M. Duan and H. J. Kimble, Phys. Rev. Lett. {\bf 90}, 253601 (2003);

% \bibitem{YuZhoZha04} B. Yu, Z.-Wei Zhou, Y. Zhang, G.-Y. Xiang, and G.-C. Guo Phys. Rev. A {\bf 70}, 014302 (2004);

 \bibitem{DiMutScu05} T. Di, A. Muthukrishnan, M. O. Scully, and M. S. Zubairy, Phys. Rev. A {\bf 71}, 062308 (2005).

\bibitem{RieHafRoo04} M. Riebe, H. H\"affner, C. F. Roos, W. H\"ansel, J. Benhelm, G. P. T. Lancaster, T. W. K\"orber, C. Becher, F. Schmidt-Kaler, D. F. V. James, and R. Blatt , Nature {\bf 429}, 734 (2004);

\bibitem{BarChiSch04} M. D. Barrett, J. Chiaverini, T. Schaetz, J. Britton, W. M. Itano, J. D. Jost, E. Knill, C. Langer, D. Liebfried, R. Ozeri, and D. J. Wineland, Nature  429, 737 (2004).


\bibitem{DuaCirZol00} L.-M. Duan, J. I. Cirac, P. Zoller, and E. S. Polzik, Phys. Rev. Lett. {\bf 85} 5643 (2000).

\bibitem{JulKozPol01} B. Julsgaard, A. Kozhekin, and E. S. Polzik, Nature {\bf 413}, 400 (2001)

\bibitem{MatKuz04} D. N. Matsukevich and A. Kuzmich, Science {\bf 306}, 22 (2004).

\bibitem{LooLadSan06} P. van Loock, T. D. Ladd, K. Sanaka, F. Yamaguchi,
Kae Nemoto, W. J. Munro, and Y. Yamamoto, Phys. Rev. Lett. {\bf 96},
240501 (2006).

\bibitem{EnkKim01} S. J. van Enk and H. J. Kimble, Phys. Rev. A 63, 023809 (2001).

\bibitem{YurMcCKla86} B. Yurke, S. L. McCall, and J. R. Klauder, Phys. Rev. A {\bf 33}, 4033 (1986).

\bibitem{HolBur93} M. J. Holland and K. Burnett, Phys. Rev. Lett. {\bf 71} 1355 (1993).

%\bibitem{MAN} M. A. Nielsen, E. Knill, and  R. Laflamme, Nature {\bf 396}, 52 (1998).

\bibitem{PezSme06} L. Pezz\'e and A. Smerzi, Phys. Rev. A {\bf 73} 011801(R) (2006).

\bibitem{HeiHorRey87} A. Heidmann, R. J. Horowicz, S. Reynaud, E. Giacobino, C. Fabre, and G. Camy, Phys. Rev. Lett. {\bf 59} 2555 (1987).

\bibitem{LanFriRon02} B. Lantz, P. Fritschel, H. Rong, E. Daw, and G. Gonz\'alez, J. Opt. Soc. Am. A {\bf 19}, 91 (2002).

\bibitem{BolItaWin96} J. J. Bollinger, Wayne M. Itano, D. J. Wineland, and D. J. Heinzen, Phys. Rev. A {\bf 54}, R4649 (1996).

\bibitem{Ger00} C. C. Gerry, Phys. Rev. A. {\bf 61}, 043811,(2000)

\bibitem{MunNemMil02} W. J. Munro,  K. Nemoto, G. J. Milburn,  S. L. Braunstein, Phys. Rev. A {\bf 66}, 023819 (2002)

\bibitem{HueMacPel97} S. F. Huelga, C. Macchiavello, T. Pellizzari, A. K. Ekert, M. B. Plenio and J. I. Cirac, Phys. Rev. Lett. {\bf 79}, 3865
(1997).

\bibitem{HooKimYe01} C J. Hood, H. J. Kimble, and J.Ye, Phys. Rev. A {\bf 64}, 033804(R) (2001)

\bibitem{RaiBruHar01} J. M. Raimond, M. Brune, and S. Haroche, Rev. Mod. Phys. {\bf 73}, 565 (2001).

\bibitem{KuhHenRem02} A. Kuhn, M. Hennrich, and G. Rempe, Phys. Rev. Lett.{\bf 89}, 067901 (2002).

\bibitem{IrvHenBou06} W. T. M. Irvine, K. Hennessy, and D. Bouwmeester,  Phys. Rev. Lett. {\bf 96}, 057405 (2006).

%\bibitem{WL} W. Lange and H. J. Kimble, Phys. Rev. A {\bf 61}, 063817 (2000).

%\bibitem{SEH} S. E. Hamann $et~al.$, Phys. Rev. Lett. {\bf 80}, 4149 (1998).

%\bibitem{WH} W. Hansel $et~al.$, Nature {\bf 413}, 466 (2001).

\bibitem{HraReh05} Z. Hradil and J. \v{R}eh\'{a}\v{c}ek, Phys. Lett. A {\bf 334}, 267 (2005).

\bibitem{PezSme05} L. Pezz\'{e} and A. Smerzi, e-print quant-ph/0508158 (2005).

\bibitem{KimDunBur05} T. Kim, J. Dunningham, and K. Burnett, Phys. Rev. A {\bf 72}, 055801 (2005).

\bibitem{ZouKimLee01} X.-B. Zou, J. Kim, and H.-W. Lee, Phys. Rev. A {\bf 63}, 065801 (2001).

\bibitem{MitLunSte04} M. W. Mitchell, J. S. Lundeen, and A. M. Steinberg, Nature (London) {\bf 429}, 161 (2004).

\bibitem{GerCam01} C. C. Gerry and R. A. Campos, Phys. Rev. A {\bf 64}, 063814 (2001).

\bibitem{LeeKokCer02} H. Lee, P. Kok, N. J. Cerf, and J. P. Dowling, Phys. Rev. A
{\bf 65}, 030101(R) (2002).

\bibitem{YurMcCKla85} B. Yurke, S. L. McCall, and J. R. Klauder,
Phys. Rev. A {\bf 33}, 4033 (1986).

\bibitem{MicJakCir03} A. Micheli, D. Jaksch, J. I. Cirac, and P. Zoller, Phys. Rev. A {\bf 67}, 013607 (2003).

\bibitem{HuaMoo06} Y. P. Huang and M. G. Moore, Phys. Rev. A {\bf 73}, 023606 (2006)

\bibitem{GreHorZei89} D. M. Greenberger, M. Horne, and A. Zeilinger, in \textit{Bell's Theorem, Quantum Theory, and Conceptions of the Universe}, edited by M. Kafatos (Kluwer, Dordrecht, 1989);

\bibitem{PanBouWei98} J.-W. Pan, D. Bouwmeester, H. Weinfurter, and A. Zeilinger, Phys. Rev. Lett. {\bf 80}, 3891 (1998).

\bibitem{DuaLukCir01} L.-M. Duan, M. D. Lukin, J. I. Cirac, and P. Zoller, Nature {\bf 414} 413 (2001).


\end{thebibliography}
\end{document}